\documentclass[aps,prc,preprint]{revtex4}
 \usepackage{amsfonts}
\usepackage{amssymb}
\usepackage{graphics}
\usepackage{graphicx}
\def\v#1{\mbox{\boldmath$#1$}}
\def\ket#1{|#1 \rangle}
\def\bra#1{\langle #1|}

\newcommand{\mc}{\multicolumn}
\newcommand{\lsim}{\mathrel{\mathop{\kern 0pt \rlap
  {\raise.2ex\hbox{$<$}}}
  \lower.9ex\hbox{\kern-.190em $\sim$}}}
\newcommand{\gsim}{\mathrel{\mathop{\kern 0pt \rlap
  {\raise.2ex\hbox{$>$}}}
  \lower.9ex\hbox{\kern-.190em $\sim$}}}

\begin{document}

\title{The Non--Mesonic Weak Decay of Double--$\Lambda$ Hypernuclei:
A Microscopic Approach}

\author{E. Bauer$^1$, G. Garbarino$^2$ and C. A. Rodr\'{\i}guez Pe\~{n}a$^3$}

\affiliation{$^1$Facultad de Ciencias Astron\'omicas y
Geof\'{\i}sicas, Universidad Nacional de La Plata and
IFLP, CONICET C. C. 67, 1900 La Plata, Argentina}

\affiliation{$^2$IIS G. Peano, I-10125 Torino, Italy}

\affiliation{$^3$Departamento de F\'{\i}sica, Universidad Nacional de La Plata,
C. C. 67, 1900 La Plata, Argentina}

\date{\today}

\begin{abstract}
The non--mesonic weak decay of double--$\Lambda$ hypernuclei is studied within a
microscopic diagrammatic approach. Besides the nucleon--induced mechanism, $\Lambda N\to nN$,
widely studied in single--$\Lambda$ hypernuclei, additional hyperon--induced mechanisms,
$\Lambda \Lambda\to \Lambda n$, $\Lambda \Lambda\to \Sigma^0 n$
and $\Lambda \Lambda\to \Sigma^-p$, are
accessible in double--$\Lambda$ hypernuclei and are investigated here.
As in previous works on single--$\Lambda$ hypernuclei,
we adopt a nuclear matter formalism extended to finite nuclei via the local density
approximation and a one--meson exchange weak transition potential
(including the ground state pseudoscalar and vector octets mesons) supplemented
by correlated and uncorrelated two--pion--exchange contributions.
The weak decay rates are evaluated for hypernuclei
in the region of the experimentally accessible light hypernuclei
$^{10}_{\Lambda\Lambda}$Be and $^{13}_{\Lambda\Lambda}$B.
Our predictions are compared with a few previous evaluations.
The rate for the $\Lambda \Lambda\to \Lambda n$ decay is dominated by
$K$--, $K^*$-- and $\eta$--exchange and turns out to be
about 2.5\% of the free $\Lambda$ decay rate, $\Gamma_{\Lambda}^{\rm free}$,
while the total rate for the
$\Lambda \Lambda\to \Sigma^0 n$ and $\Lambda \Lambda\to \Sigma^- p$ decays,
dominated by $\pi$--exchange,
amounts to about 0.25\% of $\Gamma_{\Lambda}^{\rm free}$.
The experimental measurement of these decays would be essential for the
beginning of a systematic study of the non--mesonic decay of strangeness $-2$ hypernuclei.
This field of research could also shed light on the possible existence
and nature of the $H$--dibaryon.
\end{abstract}



\maketitle

\section{Introduction}
\label{Introduction}

Strangeness nuclear physics plays an important
role in modern nuclear and hadronic physics and involves
important connections with astrophysical processes and
observables as well as with QCD. In particular,
the weak decay of $\Lambda$ hypernuclei is the only actual source of information
on strangeness--changing four--baryon weak interactions. A great variety of
theoretical and experimental studies were performed on the decay of such systems.
Let us mention the experimental and theoretical analysis
of nucleon--coincidence emission spectra and the theoretical modeling
of the decay channels within complete one--meson--exchange weak transition
potentials, which in some case have been supplemented by a two--pion--exchange mechanism.
A reasonable agreement between data and predictions have been reached
for the mesonic and non--mesonic decay rates, the $\Gamma_n/\Gamma_p$ ratio
between the neutron-- and the proton--induced decay widths,
the $\Gamma_2/\Gamma_{\rm NM}$ ratio between the two--nucleon induced
and the total non--mesonic rates,
and the intrinsic asymmetry parameter $a_\Lambda$
for the decay of polarized hypernuclei \cite{reports}. Nevertheless,
discrepancies between theory and experiment are still
present for the emission spectra involving protons \cite{BG11}.

Despite their implications on the possible existence of
dibaryon states and multi--strangeness hypernuclei and on the study of compact stars,
much less is known on strangeness $-2$ hypernuclei. Little information is available on
cascade hypernuclei, for instance on the $\Xi$--nucleus potential.
The existence of the strong $\Xi^-p \to \Lambda \Lambda$ reaction
makes $\Xi$ Hypernuclei unstable with respect to the strong interaction.
However, this conversion reaction can be exploited to produce
double--$\Lambda$ hypernuclei.

Investigations on the structure of double--$\Lambda$ hypernuclei are important to
determine the $\Lambda \Lambda$ strong interaction, which is poorly known at present.
Indeed, only a few double--$\Lambda$ hypernuclei events have been studied experimentally up
to date. In KEK experiments, $^4_{\Lambda \Lambda}$H, $^6_{\Lambda \Lambda}$He
and $^{10}_{\Lambda \Lambda}$Be have been identified, while less unambiguous
events were recorded for $^6_{\Lambda \Lambda}$He and $^{10}_{\Lambda \Lambda}$Be
in the 60's and for $^{13}_{\Lambda \Lambda}$B in the early 90's \cite{Nak10}.
The observation of the so--called NAGARA event
implies a weak and attractive $\Lambda \Lambda$ interaction,
i.e., a bond energy $\Delta B_{\Lambda \Lambda}(^6_{\Lambda \Lambda}{\rm He})\equiv
B_{\Lambda \Lambda}(^6_{\Lambda \Lambda}{\rm He})
-2B_\Lambda(^5_{\Lambda}{\rm He})= (0.67 \pm 0.17)$ MeV \cite{Tak01}.
In Ref.\cite{GM12} the authors demonstrated that this bond energy value,
which will be employed in the present work as the binding energy
between the two $\Lambda$'s,
describes well double--$\Lambda$ hypernuclear data in the mass range from 6 to 13.
%
Future experiments on strangeness $-2$ hypernuclei will be carried out at J--PARC \cite{J-PARC}
and FAIR (PANDA Collaboration) \cite{Panda}.

On the weak interaction side, double--$\Lambda$ hypernuclei offer the opportunity to
access the following $\Lambda$--induced $\Lambda$ decay channels:
$\Lambda \Lambda\to \Lambda n$, $\Lambda \Lambda\to \Sigma^- p$, $\Lambda \Lambda\to \Sigma^0 n$
(with a $\Delta S=1$ change in strangeness) and $\Lambda \Lambda\to n n$ ($\Delta S=2$).
The initial $\Lambda\Lambda$ pair is coupled to $S=0$ and $J=0$, thus only two non--mesonic decay
channels are accessible: $^1 S_0\to$$^1 S_0$ and $^1 S_0\to$$^3 P_0$ in spectroscopic notation.
No data is available on these decays, apart from
the claim for the observation of a single event at KEK \cite{Wa07}.
The experimental signature of a $\Lambda \Lambda\to \Lambda n$
decay is clear, i.e., the emission of a large momentum $\Lambda$ ($\sim 425$ MeV),
but the major problem is that these events are expected to be rather rare.
The usual neutron-- and proton--induced decays,
$\Lambda n\to nn$ and $\Lambda p\to np$, dominate over the $\Lambda$--induced ones
in double--$\Lambda$ hypernuclei.

Realistic calculation and improved measurements of the $\Lambda$--induced $\Lambda$
weak decays could also provide hints on the possible existence of the long--hunted
$H$--dibaryon. A reliable calculation is important in
the design of future experiments at J--PARC and FAIR,
where these weak processes could be unambiguously observed for the first time.

Only a few predictions are available
for such interesting strangeness--changing processes \cite{Pa01,It01,Sa03};
unfortunately, there are major disagreements among the predictions of these works,
which adopted different frameworks.
Their results are discussed in the following together with the new ones obtained here.

In this paper we present a microscopic calculation of both the $\Lambda$--
and nucleon--induced $\Lambda$ decay rates for double--$\Lambda$ hypernuclei
by using a nuclear matter formalism (the $\Lambda \Lambda\to nn$ decay
channel is not considered here since, requiring a strangeness variation of 2 units,
it is much less likely then the other $\Lambda$--induced processes); results
for finite hypernuclei in the mass range of the empirically interesting
$^{10}_{\Lambda\Lambda}$Be and $^{13}_{\Lambda\Lambda}$B systems
are reported within the local density approximation.
The same microscopic approach showed that
Pauli exchange and ground state correlation contributions are very important for a
detailed calculation of the rates, the asymmetry parameter
and the nucleon emission spectra
in the non--mesonic weak decay of $\Lambda$ hypernuclei \cite{BG09,BG10,BG11,BGPR12}.
Less pronounced effects have been reported by including the $\Delta$--baryon
resonance in the microscopic approach \cite{BG12}.

The paper is organized as follows. In Section \ref{Formalism} we present the
theoretical formalism employed for the evaluation of the decay rates. In
Section \ref{Results} the numerical results are discussed and compared with
previous calculations. Finally, in Section \ref{Conclusions} we draw our conclusions.

\section{Formalism}
\label{Formalism}

Let us start by writing
the total non--mesonic decay rate for a double--$\Lambda$ hypernucleus as:
\begin{equation}
\Gamma_{\rm NM}=\Gamma_{\rm N}+\Gamma_{\rm \Lambda}\,,
\end{equation}
where:
\begin{eqnarray}
\Gamma_{\rm N}&=&
\Gamma(\Lambda n\to nn)+\Gamma(\Lambda p\to np)\equiv\Gamma_{n}+\Gamma_{p}\, , \\
\Gamma_{\rm \Lambda}&=&\Gamma(\Lambda \Lambda\to \Lambda n)+\Gamma(\Lambda \Lambda \to \Sigma^0 n)
+\Gamma(\Lambda \Lambda\to \Sigma^- p) \\
&& \equiv\Gamma_{\Lambda n}+\Gamma_{\Sigma^o n}+\Gamma_{\Sigma^- p}\, , \nonumber
\end{eqnarray}
are the total nucleon-- and $\Lambda$--induced decay rates, respectively.
The definitions of the partial rates $\Gamma_{n}$, $\Gamma_{p}$,
$\Gamma_{\Lambda n}$, $\Gamma_{\Sigma^o n}$ and $\Gamma_{\Sigma^- p}$ are
self--explanatory. We do not consider two--baryon
induced decay mechanisms.

As in previous papers on $\Lambda$ hypernuclei,
we adopt a microscopic formalism. In this
many--body technique the calculation is performed in infinite nuclear
matter and then it is extended to finite nuclei through
the local density approximation (LDA)~\cite{Os85}.

\begin{figure}[h]
\begin{center}
    \includegraphics[width = 0.75 \textwidth]{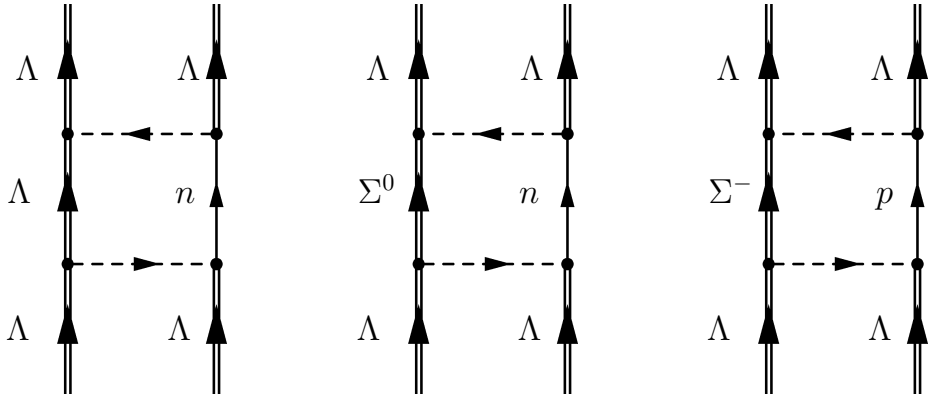}
\vskip 2mm
\caption{Goldstone diagrams for the evaluation of the $\Lambda \Lambda\to \Lambda n$,
$\Lambda \Lambda\to \Sigma^0 n$ and $\Lambda \Lambda\to \Sigma^- p$
decay rates in infinite nuclear matter.}
\label{figme1}
\end{center}
\end{figure}

The many--body contributions we consider for describing the $\Lambda \Lambda\to Y N$
processes in nuclear matter
are given by the Goldstone diagrams of Fig.\ref{figme1}.
They provide the various decay
widths through the relation $\Gamma_f = -2\, {\rm Im}\, \Sigma^{\Lambda\Lambda}_f$,
$\Sigma^{\Lambda\Lambda}_f$ being the $\Lambda\Lambda$ self--energy and
$f = \Lambda n, \Sigma^0 n$ and $\Sigma^- p$ denoting the possible final states.

Let us consider infinite nuclear matter with Fermi momentum $k_F$ and
denote the four--momenta of the initial $\Lambda$'s with
$k=(k_0,\v{k})$ and $k'=(k'_0,\v{k'})$ and the
four--momenta of the final particles by
$p_1=(p_{1 \, 0},\v{p}_1)$ (hyperon) and $p_2=(p_{2 \, 0},\v{p}_2)$ (nucleon).
In a schematic way, for the Goldstone diagrams of Fig. \ref{figme1}
one obtains the partial decay width
to the $YN$ ($=\Lambda n$, $\Sigma^0 n$ and $\Sigma^- p$) final state as follows:
\begin{equation}
\label{decwYN}
\Gamma_{Y N}(\v{k},k_{F}) = \sum_{f} \,
 |\bra{f} V^{\Lambda \Lambda \to Y N} \ket{0}_{k_{F}}|^{2}
\delta (E_{f}-E_{0})~,
\end{equation}
where $V^{\Lambda \Lambda \to Y N}$ is the weak
transition potential, $\ket{0}_{k_{F}}$ denotes the initial state with energy $E_0$
including the nuclear matter ground state and the
two $\Lambda$'s in the $1s$ level, and $\ket{f}$ the
possible final states with energy $E_f$ including nuclear matter and the $YN$ pair.
Note also that momentum conservation, i.e.,
$\v{k'}=\v{p_1}+\v{p_2}-\v{k}$, implies that only one of the initial momenta
($\v{k}$) is an independent variable once $\v{p_1}$ and $\v{p_2}$ are integrated out,
as in Eq.~(\ref{decwYN}).

The decay rates for a finite hypernucleus are obtained from the previous
partial widths via the LDA:
\begin{equation}
\label{decwYNld}
\Gamma_{Y N} =
\int d \v{k} \, |\widetilde{\psi}_{\Lambda}(\v{k})|^2
 \int d \v{r} \, |\psi_{\Lambda}(\v{r})|^2 \,
\Gamma_{Y N}(\v{k},k_{F}(r))~.
\end{equation}
This approximation (see also Appendix \ref{APPEND-rate})
consists in introducing a
local nucleon Fermi momentum $k_{F}(r) = \{3 \pi^{2} \rho(r)/2\}^{1/3}$ in terms of
the density profile $\rho(r)$ of the nuclear core and then in averaging
the partial widths over the nuclear volume. This average is
weighted by the probability per unit volume
of finding the $\Lambda$ which then transforms into the final nucleon
at a given position $\v{r}$, $|\psi_{\Lambda}(\v{r})|^2$.
A further average is performed over
the momentum distributions of the $\Lambda$, $\widetilde{\psi}_{\Lambda}(\v{k})$
(both initial $\Lambda$'s lies in the $1s_{1/2}$ single--particle state).
The calculation is performed for double--$\Lambda$ hypernuclei with mass number
$A= 10$--$13$ in order to mimic the behavior of the experimentally accessible finite
hypernuclei $^{10}_{\Lambda\Lambda}$Be and $^{13}_{\Lambda\Lambda}$B.
As in Ref.~\cite{Pa01}, for the function $\psi_{\Lambda}(\v{r})$ we use a
$1s_{1/2}$ harmonic oscillator wave--function; its
frequency $\hbar \omega=13.6$ MeV is obtained from
the fit of Ref.~\cite{Ca99} of the experimental binding energies of
$^{6}_{\Lambda\Lambda}$He, $^{10}_{\Lambda\Lambda}$Be and $^{13}_{\Lambda\Lambda}$B.
The energies of the initial $\Lambda$ with momentum $\v{k}$ is given by
$k_{0}=m_\Lambda+\v{k}^2/(2 m_\Lambda)+V_{\Lambda}$,
where for the binding term we adopt the value $V_\Lambda=-\hbar \omega=-13.6$ MeV.

Before we give explicit expressions for the decay widths in nuclear matter,
it is convenient to show the general form of the weak transition potential.
The standard weak, strangeness--changing transition potential for the
$\Lambda \Lambda \to Y N$ processes can be written as:
\begin{equation}
\label{intergral}
V^{\Lambda \Lambda \to Y N} (q) = \sum_{\tau=0,1} {\cal O}_\tau
{V}_\tau(q),~~~~~ {\cal O}_\tau=\left\{
\begin{array}{c}1~~\mbox{for}~~\tau=0\\
  \v{\tau}_1 \cdot \v{\tau}_2~~\mbox{for}~~\tau=1
\end{array}\right.,
\end{equation}
where
\begin{eqnarray}
\label{allmespot}
{V}_{\tau}(q)
=   (G_F m_{\pi}^2)  && \{
S_{\tau}(q)  \; \v{\sigma_1 \cdot \hat{q}} +
S'_{\tau}(q)  \; \v{\sigma_2 \cdot \hat{q}} +
P_{L, \tau}(q)  \v{\sigma_1 \cdot \hat{q}} \; \v{\sigma_2 \cdot
\hat{q}} \nonumber \\
& & +P_{C, \tau}(q)  +    P_{T, \tau}(q)  (\v{\sigma_1 \times \hat{q}})
\cdot  (\v{\sigma_2 \times \hat{q}})  \nonumber \\
& &  +i S_{V, \tau}(q)
\v{(\sigma_1 \times \sigma_2) \cdot \hat{q}} \}\,.
\end{eqnarray}
The functions $S_{\tau}(q)$, $S'_{\tau}(q)$, $P_{L, \tau}(q)$,
$P_{C,\tau}(q)$, $P_{T, \tau}(q)$ and $S_{V, \tau}(q)$
contain baryon--baryon short range correlations and vertex form factors
and are taken from the Appendix B of Ref.~\cite{ba03},
with the modifications concerning the baryon coupling constants discussed in the
Appendix~\ref{APPEND} of the present paper.
The values $\tau=0,1$ stand for the isoscalar
and isovector parts of the interaction, respectively.

\begin{figure}[h]
\begin{center}
    \includegraphics[width = 0.7\textwidth]{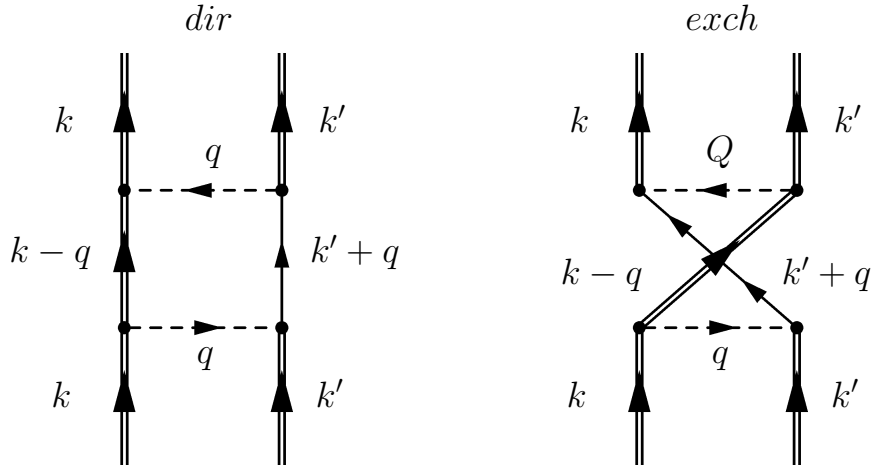}
\vskip 2mm
\caption{Direct and exchange Goldstone diagrams for the
$\Lambda \Lambda \to \Lambda n$ decay.}
\label{figme2}
\end{center}
\end{figure}

To enforce antisymmetrization, for each one of the contributions of Fig.~\ref{figme1}
we also consider the corresponding exchange contribution.
In Fig.~\ref{figme2} we give the direct and exchange diagrams
for $\Lambda \Lambda \to \Lambda n$.
Through the standard rules for Goldstone diagrams one writes down
the explicit expression for these contributions.
After performing the summations over spin and isospin together with the
energy--integration one obtains the antisymmetrized decay rate:
\begin{eqnarray}
\label{gamdLn}
\Gamma_{\Lambda n}(\v{k},k_{F}) & =
& \pi \, (G_F m_{\pi}^2)^2  \int \frac{d^{3} p_1}{(2 \pi)^{3}} \int \frac{d^{3} p_2}{(2 \pi)^{3}}
\;  (2 \, {\cal W}^{dir}_{0}(q)- {\cal W}^{exch}_{0}(q,Q))\nonumber \\
& &  \times\theta(|\v{p}_2|-k_F) \;
\delta(k_0+k'_0-E_{\Lambda}(p_1)-E_{n}(p_2))\, ,\nonumber
\end{eqnarray}
where $E_{\Lambda}$ ($E_{n}$) is the total $\Lambda$ (neutron) energy, while
$q=k-p_1$ and $Q=p_2-k$.
For the direct term, the momentum matrix--element of the interaction turns out to be:
\begin{equation}
\label{vdir}
{\cal W}^{dir}_{0}(q)  =  \{
S^{2}_{0}(q) + S'^{2}_{0}(q)
 + P^{2}_{L, 0}(q)  +  P^{2}_{C, 0}(q)
 + 2  \, P^{2}_{T, 0}(q) + 2 \, S^{2}_{V,0}(q) \}\,,
\end{equation}
while for the exchange term we have:
\begin{eqnarray}
\label{vexc}
{\cal W}_{exch}(q,Q) & = &
(\hat{\v{q}} \cdot \hat{\v{Q}})  \textsf{S}_{0}(q,Q)
 + ( 2 (\hat{\v{q}} \cdot \hat{\v{Q}})^2 - 1)  P_{L,
0}(q)  P_{L, 0}(Q)  \nonumber \\
& & + 2 ((\hat{\v{q}} \cdot \hat{\v{Q}})^2 - 1)   P_{T,
0}(q)  P_{T, 0}(Q) \nonumber \\
& &  -2 (\hat{\v{q}} \cdot \hat{\v{Q}})^2 (P_{L,
0}(q) P_{T, 0}(Q) + P_{L, 0}(Q) P_{T,
0}(q)) \nonumber \\
 & & + P_{C, 0}(q) P_{C, 0}(Q) +
P_{C, 0}(q) P_{L, 0}(Q) + P_{C, 0}(Q) P_{L, 0}(q) \nonumber \\
& & + 2 (P_{C, 0}(q) P_{T, 0}(Q) + P_{C, 0}(Q) P_{T,0}(q))\,,
\end{eqnarray}
where we have defined:
\begin{eqnarray}
\label{sexc}
\textsf{S}_{0}(q,Q) & = &
(S_{0}(q) + S'_{0}(q))(S_{0}(Q)
+ S'_{0}(Q))
\nonumber \\
& & - 2 (S_{0}(q) S_{V, \, 0}(Q) +
S_{V, \, 0}(q) S_{0}(Q)) \nonumber \\
& & + 2 (S'_{0}(q) S_{V, \, 0}(Q) +
S_{V, \, 0}(q) S'_{0}(Q))\,.
\end{eqnarray}
Note from Eqs.~(\ref{vdir})--(\ref{sexc})
that, being the $\Lambda\Lambda\to \Lambda n$ weak potential
of isoscalar nature, we have fixed $\tau=0$ in Eqs.~(\ref{intergral})
and (\ref{allmespot}).
In Appendix~\ref{APPEND} we present explicit expressions for the
$\Lambda\Lambda \to \Sigma^{0} n$ and $\Lambda\Lambda \to \Sigma^{-} p$
decay rates.

We assume the
$\Delta I = 1/2$ rule on the isospin change to be valid for all weak transition,
although it is phenomenologically justified only for the $\Lambda N \pi$
weak free vertex.
Thus, by neglecting the small mass difference between $\Sigma^0$ and $\Sigma^-$ one
simply obtains that
the rates for decays into $\Sigma^0 n$ and $\Sigma^- p$ states are simply related by:
\begin{equation}
\label{rule}
\frac{\Gamma_{\Sigma^- p}}{\Gamma_{\Sigma^0 n}}=2\, ,
\end{equation}
and it is sufficient to calculate the decay rates $\Gamma_{\Lambda n}$ and
$\Gamma_{\Sigma^0 n}$.

We adopt a meson--exchange description of the weak transition potential including
$\pi$, $\eta$, $K$, $\rho$, $\omega$ and $K^*$ mesons (these contribute
to the one--meson--exchange part, denote by OME in the following)
together with a two--pion--exchange mechanism (TPE).
The latter has been obtained from the $\Lambda N\to \Lambda N$
scalar--isoscalar two--pion--exchange strong interaction potential
derived in Ref.~\cite{Sa07} by a chiral unitary approach and
consists in both an uncorrelated and a correlated part. The present work is the
first one to include the TPE mechanism.
Since isospin is conserved in strong vertex,
the $\Lambda\Lambda \to \Lambda n$ decay process has isoscalar character and only
the $\eta$, $K$, $\omega$, $K^*$ exchange and TPE contribute, while
for $\Lambda\Lambda \to \Sigma^0 n$ isoscalar transitions are prohibited and
the contributing mesons are $\pi$, $K$, $\rho$ and $K^*$.
At the OME level one naively expects the $\Lambda \Lambda\to \Lambda n$ decay
($\Lambda \Lambda\to \Sigma^- p$, $\Lambda \Lambda\to \Sigma^0 n$ decays) to be
dominated by $K$--exchange ($\pi$--exchange). In particular, from the
$\Lambda \Lambda \to \Lambda n$ ($\Lambda \Lambda \to \Sigma^0 n$)
channel one could obtain information on
the $\Lambda \Lambda K$ ($\Lambda \Sigma K$) vertex; these vertices are important to
constrain $SU(3)$ chiral perturbation theory \cite{Pa01}.

Analyzes of $\Sigma$ formation spectra in the $(K^-,\pi^\pm)$ and
$(\pi^+,K^+)$ reactions showed that the $\Sigma$--nucleus potential
has a substantial isospin--dependence and,
with the exception of very light systems (the only quasibound state of a $\Sigma$
in a nucleus has been observed in $^4_\Sigma$He),
is repulsive: $V_\Sigma \sim +(10-50)$ MeV
at normal nuclear density. In the present calculation we adopt the
value $V_\Sigma =+30$ MeV.

\section{Results}
\label{Results}

The calculations refer to the mass range corresponding to the experimentally
accessible $^{10}_{\Lambda\Lambda}$Be and $^{13}_{\Lambda\Lambda}$B hypernuclei. Practically,
the calculations are performed with $A=N+Z+2=12$
and an equal number of neutrons and protons, $N=Z=5$.
We verified that the numerical results does not change appreciably by changing $A$ by one
or two units: we will refer to them as the results for $A\sim 12$ double--$\Lambda$
hypernuclei.

In Table \ref{results-partial} we give our results
for the $\Lambda\Lambda\to \Lambda n$,
$\Lambda\Lambda\to \Sigma^0 n$ and $\Lambda\Lambda\to \Sigma^- p$ weak decay widths.
Predictions are given for the individual meson exchanges
and for the most relevant combinations among them.
\begin{table}[h]
\begin{center}
\caption{Results for the $\Lambda\Lambda\to \Lambda n$,
$\Lambda\Lambda\to \Sigma^0 n$ and $\Lambda\Lambda\to \Sigma^- p$ weak decay widths in
$A\sim 12$ double--$\Lambda$ hypernuclei are given as a
percentage of the free $\Lambda$ decay rate. Predictions are given
for the individual contributing mesons and for the most relevant meson combinations.}
\label{results-partial}
\begin{tabular}{l c c c c} \hline\hline
\mc {1}{c}{Model and Ref.} &
\mc {1}{c}{$~~~~~\Gamma_{\Lambda n}~~~~~$} &
\mc {1}{c}{$~~~~~\Gamma_{\Sigma^0 n}~~~~~$} &
\mc {1}{c}{$~~~~~\Gamma_{\Sigma^- p}~~~~~$} \\ \hline
$\pi$                                     & -- &
                                          $0.070$ &
                                          $0.140$ & \\
$K$                                       & $1.73$ &
                                          $0.001$ &
                                          $0.002$ &  \\
$\eta$                                    & $0.35$ &
                                           -- &
                                           -- &  \\
$\rho$                                    & -- &
                                          $0.001$ &
                                          $0.002$ &  \\
$K^*$                                     & $0.84$ &
                                          $0.002$ &
                                          $0.004$ &  \\
$\omega$                                  & $0.01$ &
                                           -- &
                                           -- &  \\
TPE                                       & $0.002$ &
                                           -- &
                                           -- &  \\
$\pi+K+K^*$                               & $4.14$ &
                                          $0.081$  &
                                          $0.162$  & \\
$\pi+K+K^*+\eta$                        & $2.57$ &
                                          $0.081$ &
                                          $0.162$  & \\
All                                        & $2.48$ &
                                             $0.084$
                                           & $0.168$ \\
\hline\hline
\end{tabular}
\end{center}
\end{table}
Note that the results for $\Gamma_{\Sigma^- p}$ are obtained as
$\Gamma_{\Sigma^- p}=2 \Gamma_{\Sigma^0 n}$ since only $\Delta I = 1/2$
transitions are considered here.
As anticipated, the rate $\Gamma_{\Lambda n}$ ($\Gamma_{\Sigma^0 n}$) has no
contribution from isovector (isoscalar) mesons.

In the OME sector the rate $\Gamma_{\Lambda n}$ receives major contributions
by $K$-- and $K^*$--exchange. The $\eta$ contribution is smaller but non--negligible.
Instead, both the $\omega$--exchange and the TPE contributions are negligible;
the TPE provides the smallest contribution.
The addition of $K$-- and $K^*$--exchange provides a decay rate which is
about 65\% larger than the complete result for $\Gamma_{\Lambda n}$
because of a constructive interference
between the two meson contributions.
However, the further addition of the $\eta$ meson, due to a destructive
interference, lowers the decay rate to be only 4\% larger than the complete
result.

The rates $\Gamma_{\Sigma^0 n}$ and $\Gamma_{\Sigma^- p}$ are much smaller
than $\Gamma_{\Lambda n}$ and, as expected, are dominated by $\pi$--exchange.
Much smaller single contributions originate from $K$--, $K^*$-- and $\rho$--exchange.
However, the combined effect of these mesons is to increase the rates by about 20\%
thanks to constructive interference effects.
From the kinematics point of view,
mesons heavier than the pion are expected to contribute less to the
rates $\Gamma_{\Sigma^0 n}$ and $\Gamma_{\Sigma^- p}$ than to the rate
$\Gamma_{\Lambda n}$ since the $\Lambda \Lambda \to \Lambda n$ process
is characterized by larger momentum transfers than the
$\Lambda \Lambda \to \Sigma^0 n$ and $\Lambda \Lambda \to \Sigma^- p$
processes. This is confirmed by the results of Table \ref{results-partial}:
the $\Gamma_{\Lambda n}$ rate receives substantial
contributions from $K$--, $K^*$-- and $\eta$--exchange,
while $\Gamma_{\Sigma^0 n}$ is dominated by $\pi$--exchange.

Before comparing the above predictions with those obtained in previous calculations,
in Table \ref{results-n} we present our results for
the nucleon--induced non--mesonic decay rates together with
determinations from Refs.~\cite{Pa01,It01}.
\begin{table}[h]
\begin{center}
\caption{Predictions for the nucleon--induced non--mesonic weak decay rates
for $A\sim 12$ double--$\Lambda$ hypernuclei.
The results of the present work are given together with
previous ones available for $^{6}_{\Lambda\Lambda}$He \cite{Pa01,It01}.
The decay rates are in units of the free $\Lambda$ decay width.}
\label{results-n}
\begin{tabular}{l c c c c} \hline\hline
\mc {1}{c}{Model and Ref.} &
\mc {1}{c}{$~~~~~\Gamma_n~~~~~$} &
\mc {1}{c}{$~~~~~\Gamma_p~~~~~$} &
\mc {1}{c}{$~~~~~\Gamma_n/\Gamma_p~~~~~$} &
\mc {1}{c}{$~~~~~\Gamma_{\rm N}=\Gamma_n+\Gamma_p~~~~~$} \\ \hline
This Work ($A\sim 12$)                           & 0.48   & 1.12  & 0.43  & 1.60 \\
OME ($^{6}_{\Lambda\Lambda}$He)
    \cite{Pa01}                                  & 0.30   & 0.66  & 0.46  & 0.96 \\
$\pi+2\pi/\rho+2\pi/\sigma$ ($^{6}_{\Lambda\Lambda}$He)
                                     \cite{It01} & 0.295  & 0.441 & 0.669 & 0.736 \\
\hline\hline
\end{tabular}
\end{center}
\end{table}
Our predictions for $\Gamma_n$ and $\Gamma_p$ in $A\sim 12$ double--$\Lambda$
hypernuclei are larger than previously obtained for
$^{6}_{\Lambda\Lambda}$He; indeed, it is well established that, in single--$\Lambda$
hypernuclei, the values of the
$\Lambda N\to nN$ rates are increasing as a function of $A$
and saturate for $A\sim 20$.
One expects the neutron-- and proton--induced rates for a double--$\Lambda$ hypernucleus
to be larger than twice the corresponding
rates for a single--$\Lambda$ hypernucleus with one unit less mass number:
$\Gamma_{\rm N}(^A_{\Lambda\Lambda}Z) > 2\,\Gamma_{\rm N}(^{A-1}_\Lambda Z)$.
Apart from the fact that a double--$\Lambda$ hypernucleus has twice the number
of $\Lambda$'s than a single--$\Lambda$ hypernucleus, one has to consider that
the binding energy of a $\Lambda$ is larger in $^A_{\Lambda\Lambda}Z$
than in $^{A-1}_\Lambda Z$. This is well confirmed experimentally
by binding data on $^6_{\Lambda\Lambda}$He and $^5_\Lambda$He. The same
behavior is expected in our mass range~\footnote{Note that available experimental
values of the harmonic oscillator parameter $\hbar \omega$ (obtained as the energy
separation between the $s$ and $p$ $\Lambda$--levels) are $-10.8$ MeV
for $^{12}_\Lambda$C and $-13.6$ MeV for $A=6$--$13$ double--$\Lambda$ hypernuclei.},
although for increasing $A$ the $\Lambda$
binding energies for double--$\Lambda$ and single--$\Lambda$ hypernuclei should converge
towards a common value. Our results confirm the described behavior:
the total nucleon--induced non--mesonic decay rate obtained for an $A= 12$
double--$\Lambda$ hypernucleus,
$\Gamma_{\rm N}=1.60\, \Gamma_\Lambda^{\rm free}$, is
about 5\% larger than twice the same rate we obtain within the same
framework and weak potential model for $^{11}_\Lambda$B,
$\Gamma_{\rm N}(^{11}_\Lambda{\rm B})=0.76\, \Gamma_\Lambda^{\rm free}$.

In Table \ref{results-h} our final results for the $\Lambda\Lambda \to \Lambda n$,
$\Lambda\Lambda \to \Sigma^0 n$ and $\Lambda\Lambda \to \Sigma^- p$ decay rates
in $A\sim 12$ double--$\Lambda$ hypernuclei
are given together with existing calculations for $^{6}_{\Lambda\Lambda}$He
and $^{10}_{\Lambda\Lambda}$Be \cite{Pa01,It01,Sa03}.
\begin{table}[h]
\begin{center}
\caption{Predictions for the $\Lambda\Lambda\to \Lambda n$,
$\Lambda\Lambda\to \Sigma^0 n$ and $\Lambda\Lambda\to \Sigma^- p$  weak decay rates for
$A\sim 12$ double--$\Lambda$ hypernuclei of the present work and
for $^{6}_{\Lambda\Lambda}$He and $^{10}_{\Lambda\Lambda}$Be from previous works.
The decay rates are in units of $10^{-2}\, \Gamma_\Lambda^{\rm free}$,
$\Gamma_\Lambda^{\rm free}$ being the free $\Lambda$ decay width.}
\label{results-h}
\begin{tabular}{l c c c c} \hline\hline
\mc {1}{c}{Model and Ref.} &
\mc {1}{c}{$~~~~~\Gamma_{\Lambda n}~~~~~$} &
\mc {1}{c}{$~~~~~\Gamma_{\Sigma^0 n}~~~~~$} &
\mc {1}{c}{$~~~~~\Gamma_{\Sigma^- p}~~~~~$} \\ \hline
This Work ($A\sim 12$)                    & $2.48$  &
                                          $0.08$ &
                                          $0.17$  \\
OME ($^{6}_{\Lambda\Lambda}$He) \cite{Pa01}
                                        & $3.6$  &
                                          $0.13$ &
                                          $0.26$   \\
$\pi+K+\omega+2\pi/\rho+2\pi/\sigma$ ($^{6}_{\Lambda\Lambda}$He) \cite{It01}
                                          & $5.3$ &
                                          $0.10$ &
                                          $0.20$   \\
$\pi+K+\omega+2\pi/\rho+2\pi/\sigma$ ($^{10}_{\Lambda\Lambda}$Be) \cite{It01}
                                          & $3.4$ &
                                          $0.07$ &
                                          $0.13$   \\
$\pi+K$ ($^{6}_{\Lambda\Lambda}$He) \cite{Sa03}
                                        & $0.03$ &
                                          $0.51$ &
                                          $1.00$   \\
$\pi+K+$DQ ($^{6}_{\Lambda\Lambda}$He) \cite{Sa03}
                                        & $0.24$ &
                                          $0.65$ &
                                          $0.85$   \\
\hline\hline
\end{tabular}
\end{center}
\end{table}

Our calculation is easily comparable with the finite nucleus
(single--particle shell model) OME calculation
of Ref.~\cite{Pa01} since TPE turned out to give a negligible contribution in the
present calculation and the OME models employed in both works
have the same pseudoscalar and vector mesons content. Our predictions for
$\Gamma_{\Lambda n}$, $\Gamma_{\Sigma^0 n}$ and $\Gamma_{\Sigma^- p}$ are smaller,
by 30--40\%, than the ones of the finite nucleus calculation.
We think this is mainly due to the fact that in Ref.~\cite{Pa01} a lighter
hypernuclueus, $^6_{\Lambda\Lambda}$He, was considered.
Indeed, we proved numerically that the $\Lambda$--induced $\Lambda$
decay rate $\Gamma_{\rm \Lambda}=\Gamma_{\Lambda n}+\Gamma_{\Sigma^0 n}+\Gamma_{\Sigma^- p}$
decreases for increasing mass number $A$:
a decrease of 2\% in the rate $\Gamma_{\Lambda n}$ is obtained
if the calculation is performed with $A=10$ instead of $A=12$
(note that our LDA calculation cannot be extended to small mass numbers as $A=6$).
The results of Ref.~\cite{It01} of Table \ref{results-h} also
corroborates this behavior. Note instead that the
nucleon--induced $\Lambda$ decay rate $\Gamma_{\rm N}=\Gamma_n+\Gamma_p$
(for both single-- and double--$\Lambda$ hypernuclei) increases for increasing $A$.
The different behavior of $\Gamma_{\rm N}$ and $\Gamma_{\rm \Lambda}$ as
a function of $A$ is easily explained as follows.
On the one hand, the rate $\Gamma_{\rm N}$ increases and
then saturates with $A$ since it somehow measures the number of nucleons which
can interact with the $\Lambda$, i.e., the nucleons which can induce a $\Lambda N\to nN$
decay.  On the other hand, for increasing $A$ the average distance
between two $\Lambda$'s in a double--$\Lambda$ hypernucleus increases and
thus the rate $\Gamma_{\rm \Lambda}$ decreases.
Our $\Lambda$--induced predictions exhibit a similar behavior of the ones
of Ref.~\cite{Pa01}, which also enforced
the $\Delta I=1/2$ rule: the ratio $\Gamma_{\Lambda n}/\Gamma_{\Sigma^0 n}$
is about 28 in the finite nucleus approach, while in the present work:
\begin{eqnarray}
\frac{\Gamma_{\Lambda n}}{\Gamma_{\Sigma^0 n}}\sim 30 \,.
\end{eqnarray}

Another ratio between decay rates deserves to be considered: it involves the
neutron--induced rate $\Gamma_n$ and the $\Lambda$--induced rate $\Gamma_{\Lambda n}$.
One expect the $\Gamma_n/\Gamma_{\Lambda n}$ ratio to be driven by the
number of $\Lambda n$ pairs in the hypernucleus, i.e., by the number of
neutrons $N_n$ that can induce the non--mesonic decay. In a naive picture,
$\Gamma_n/\Gamma_{\Lambda n}$ is proportional to $N_n$. We obtain:
\begin{eqnarray}
\frac{\Gamma_{n}}{\Gamma_{\Lambda n}}\sim 19.4 \,,
\end{eqnarray}
while in the finite nucleus approach of Ref.~\cite{Pa01}
$\Gamma_n/\Gamma_{\Lambda n}\sim 8.3$.
The different results are mainly due to the different neutron numbers in the
two calculations, $N_n=5$ in the present calculation and $N_n=2$ in Ref.~\cite{Pa01}:
indeed, $(\Gamma_n/\Gamma_{\Lambda n})_{N_n=5}/(\Gamma_n/\Gamma_{\Lambda n})_{N_n=2} \sim 2.3$,
while the corresponding ratio between the neutron numbers is $5/2=2.5$.

In Ref.~\cite{It01}, a phenomenological, correlated two--pion--exchange
($2\pi/\sigma+2\pi/\rho$) mechanism was added
to a $\pi+K+\omega$--exchange model for a finite nucleus calculation for
$^6_{\Lambda \Lambda}$He and $^{10}_{\Lambda \Lambda}$Be. The authors found an
improvement in the calculation of the $\Gamma_n/\Gamma_p$ ratio for single--$\Lambda$
hypernuclei by including the $2\pi/\sigma$ and $2\pi/\rho$ potentials \cite{It02}
together with $K$--exchange \cite{It01}.
We note that in Ref.~\cite{It01} the same $\Lambda$ wave function previously adopted
for $^{5}_{\Lambda}$He was used for $^{6}_{\Lambda\Lambda}$He,
despite, as explained above, a $\Lambda$ is more bound in
$^{6}_{\Lambda\Lambda}$He than in $^{5}_{\Lambda}$He. This assumption leads to an
underestimation of the $\Gamma_n$ and $\Gamma_p$ decay rates reported in
Table \ref{results-n} for $^{6}_{\Lambda\Lambda}$He.
Concerning double--$\Lambda$ hypernuclei, in the same paper the wave function of
$^6_{\Lambda\Lambda}$He ($^{10}_{\Lambda\Lambda}$Be) was described
by an $\alpha+\Lambda+\Lambda$ three--body cluster model
($\alpha+\alpha+\Lambda+\Lambda$ four--body cluster model).
Although the final results for $^{10}_{\Lambda\Lambda}$Be are not very different from
ours, a dominant contribution from $2\pi/\sigma$--exchange to the
$\Lambda\Lambda\to \Lambda n$ decay rate is obtained; this
behavior is not confirmed by the chiral unitary approach based
TPE mechanism adopted in the present study. The lack of details from
Ref.~\cite{It01} does not allow us to understand the origin of such a discrepancy.
The ratio $\Gamma_{\Lambda n}/\Gamma_{\Sigma^0 n}$ is about 53 (49) for
$^{6}_{\Lambda\Lambda}$He ($^{10}_{\Lambda\Lambda}$Be); both results are larger
by about 80\% than what found in the present paper and in Ref.~\cite{Pa01}.
Furthermore, the ratio $\Gamma_n/\Gamma_{\Lambda n}$ is about 5.6
for $^{6}_{\Lambda\Lambda}$He, i.e.,
about 30\% less that found in the finite nucleus calculation of Ref.~\cite{Pa01}
for the same hypernucleus.

In Ref.~\cite{Sa03} an hybrid quark--meson approach is instead adopted, which includes $\pi$--
and $K$--exchange at long and medium distances and a direct quark mechanism
(basically, a valence quark picture of baryons based on an effective four--quark weak
Hamiltonian) to account for the short--range part of the
processes. The direct quark mechanism provides a large contribution to the
$\Gamma_{\Lambda n}$, $\Gamma_{\Sigma^0 n}$ and $\Gamma_{\Sigma^- p}$ decay rates
and strongly violates the isospin rule (\ref{rule}) (see the results in Table \ref{results-h}).
We note that the $\pi+K$ calculation
provides $\Gamma^K_{\Lambda n}/\Gamma^\pi_{\Sigma^0 n}=0.06$, in strong disagreement
with the other calculations of Table \ref{results-h}. We note that a simple evaluation
in terms of the weak and strong coupling constants involved in the
$\Lambda \Lambda \to \Lambda n$ decay mediated by the $K$ meson and
the $\Lambda \Lambda \to \Sigma^0 n$ decay mediated by the $\pi$ meson
indicates that the ratio $\Gamma^K_{\Lambda n}/\Gamma^\pi_{\Sigma^0 n}$
(which is a good approximation of the ratio $\Gamma_{\Lambda n}/\Gamma_{\Sigma^0 n}$;
see the results of Table \ref{results-partial}) has to be larger than 1.
When compared with the results of the present paper and of Ref.~\cite{Pa01},
the very small value of $\Gamma^K_{\Lambda n}/\Gamma^\pi_{\Sigma^0 n}$ originates
from a `very small' $K$--exchange (`large' $\pi$--exchange) contribution to
the $\Lambda \Lambda \to \Lambda n$ ($\Lambda \Lambda \to \Sigma^0 n$) channel.
We point out the strong disagreement concerning $K$--exchange:
$\Gamma^K_{\Lambda n}/(10^{-2}\Gamma^{\rm free}_\Lambda)$ is 0.03 in
the hybrid quark--meson approach, while it is 1.7 (2.7) in the present approach
(in the finite nucleus calculation of Ref.~\cite{Pa01}).
For the complete calculation, the hybrid quark--meson approach provides
$\Gamma_{\Lambda n}/\Gamma_{\Sigma^0 n}\sim 0.37$.

As mentioned, no data is available on $\Lambda$--induced $\Lambda$ decays, apart from
the claim \cite{Wa07} for the observation of a single event in the KEK hybrid--emulsion experiment
which led to the observation of the so--called NAGARA event concerning the observation
of the $^6_{\Lambda\Lambda}$He hypernucleus. The authors interpreted this event as
a weak decay of an unknown strangeness $-2$ system into a $\Sigma^- p$ pair.
This result is difficult to interpret since the KEK experimental branching
ratio (BR) for this process is of the order of $10^{-2}$ while for the
$\Lambda\Lambda\to \Sigma^- p$ decay in a double--$\Lambda$ hypernucleus
the BR is evaluated to be of the order of $10^{-3}$ in the present work as well
as in the previous determinations of Refs.~\cite{Pa01,It01}.
As done in Ref.~\cite{Wa07}, one could also speculate
that the observed event corresponds to a decay of an $H$--dibaryon.
As far as we know, there is only a dated calculation \cite{Do86} concerning the
$H\to \Sigma^- p$ process, which, adapted to the case of a double--$\Lambda$
hypernucleus, provides a BR of the order of $10^{-2}$.
Future measurements will be essential not only to establish the $\Lambda$--induced
$\Lambda$ weak decays studied here but also in order
to clarify the question of the existence and nature
of the $H$--dibaryon and eventually to establish its role
in defining the properties of double--$\Lambda$ hypernuclei.

\section{Conclusions}
\label{Conclusions}

A microscopic diagrammatic approach is used to evaluate the nucleon-- and
$\Lambda$--induced $\Lambda$ decay in double--$\Lambda$ hypernuclei.
The calculation is performed in nuclear
matter and then extended to finite hypernuclei with mass numbers $A\sim 12$
($^{10}_{\Lambda\Lambda}$Be and $^{13}_{\Lambda\Lambda}$B are experimentally
accessible cases) by means of the local density approximation.
The present approach is the first one which takes into account the full
one--meson--exchange weak transition potential together with a
two--pion--exchange contribution. The one--meson--exchange
potential contains the mesons of the ground state pseudoscalar and vector octets,
while the two--pion--exchange potential
includes correlated and uncorrelated contributions and is obtained from
the chiral unitary approach of Ref.~\cite{Sa07}.
Such a complete potential model proved to be of crucial importance in
consistently explaining the whole set of decay data on single--$\Lambda$
hypernuclei \cite{reports}.

We confirm that the neutron-- and proton--induced decay rates for the
hypernucleus $^A_{\Lambda\Lambda}Z$ with $A\sim 12$
turn out to be larger (by about 5\%)
than twice the corresponding rates for the single--$\Lambda$
hypernucleus $^{A-1}_{\Lambda}Z$; data indicates that the binding energy of a $\Lambda$
is indeed larger in  $^A_{\Lambda\Lambda}Z$ than in $^{A-1}_{\Lambda}Z$.

The two--pion--exchange mechanism turns out to
provide a negligible contribution to the $\Lambda \Lambda \to \Lambda n$
non--mesonic decay of double--$\Lambda$ hypernucleus. The rate
$\Gamma_{\Lambda n}$ receives the major contributions from $K$-- and $K^*$--exchange
(however, the $\eta$ meson cannot be neglected).
The rates $\Gamma_{\Sigma^0 n}$ and $\Gamma_{\Sigma^- p}$, which are much smaller than
$\Gamma_{\Lambda n}$ ($\Gamma_{\Lambda n}/\Gamma_{\Sigma^0 n}=29$ and
$\Gamma_{\Sigma^- p}/\Gamma_{\Sigma^0 n}=2$ in virtue of the
$\Delta I=1/2$ isospin rule),
are dominated by $\pi$--exchange.

The total $\Lambda$--induced decay rate,
$\Gamma_{\rm \Lambda}=\Gamma_{\Lambda n}+\Gamma_{\Sigma^0 n}+\Gamma_{\Sigma^- p}$,
amounts to about 1.7\% of the total non--mesonic rate,
$\Gamma_{\rm NM}=\Gamma_{n}+\Gamma_{p}+\Gamma_{\rm \Lambda}$.
We also find that the rate $\Gamma_{\rm \Lambda}$ decreases as
the hypernuclear mass number $A$ increases since  the average distance
between two $\Lambda$ in a double--$\Lambda$ hypernucleus is an increasing function
of $A$.

Our final results for $\Gamma_{\Lambda n}$,
$\Gamma_{\Sigma^0 n}$ and $\Gamma_{\Sigma^- p}$
are in fairly good agreement with the ones of Refs.~\cite{Pa01,It01}
and in strong disagreement with those of Ref.~\cite{Sa03}.

We hope the present work may contribute to the start of
a systematic investigation on the non--mesonic weak decays of double--$\Lambda$
hypernuclei. No reliable experimental evidence of interesting processes
such as $\Lambda \Lambda \to \Lambda n$,
$\Lambda \Lambda \to \Sigma^0 n$ and $\Lambda \Lambda \to \Sigma^- p$ is available
at present.
Future measurements will also be essential to clarify the question of the existence
and nature of the $H$--dibaryon and eventually to establish
its interplay and/or mixing with the $\Lambda\Lambda$ pair in determining
the structure and weak decays properties of double--$\Lambda$ hypernuclei.
New experimental programs at J--PARC and FAIR should thus be strongly supported.

\appendix
\section{}
\label{APPEND-rate}

We present here the formal derivation of Eq.~(\ref{decwYNld})
which is used to calculate the decay rates in the local density approximation (LDA).
Let us start by introducing the $\Lambda$ pair wave function in coordinate space,
$\psi_{\Lambda\Lambda}(\v{r},\v{r'})$.
In a double--$\Lambda$ hypernucleus both hyperons are paired in the lowest energy
single--particle state $1s$.
In the independent--particle approximation, $\psi_{\Lambda\Lambda}(\v{r},\v{r'})$
is simply factorized in terms of the individual $\Lambda$ wave functions
$\psi_\Lambda(\v{r})$ and $\psi_\Lambda(\v{r'})$ associated to the same energy eigenvalue:
\begin{equation}
\label{LDA0}
\psi_{\Lambda\Lambda}(\v{r},\v{r'}) = \psi_\Lambda(\v{r}) \psi_\Lambda(\v{r'})~.
\end{equation}

Let us denote with $\v{k}$ and $\v{k'}$ ($\v{p_1}$ and $\v{p_2}$)
the momenta of the initial $\Lambda$'s (final hyperon and nucleon)
for the $\Lambda\Lambda\to YN$ decay.
In the LDA one introduces the following rate
for such a decay:
\begin{equation}
\label{LDA1}
\Gamma_{Y N}(\v{k})= \int d\v{r} \int d\v{r'}
|\psi_{\Lambda\Lambda}(\v{r},\v{r'})|^2 \Gamma_{Y N}(\v{k},\v{r},\v{r'})~,
\end{equation}
$\v{k}$ being the momentum of one of the initial $\Lambda$'s.
The final momenta $\v{p_1}$ and $\v{p_2}$ are integrated out to obtain
$\Gamma_{Y N}(\v{k},\v{r},\v{r'})$.
Note also that momentum conservation, i.e.,
$\v{k'}=\v{p_1}+\v{p_2}-\v{k}$, implies that only one of the initial momenta
($\v{k}$) is an independent variable once $\v{p_1}$ and $\v{p_2}$ are integrated out.
This is the reason why the integrand in Eq.~(\ref{LDA1}) is independent of
$\v{k'}$.

The rates for finite hypernuclei are thus obtained through the relation:
\begin{equation}
\label{LDA2}
\Gamma_{Y N} =
\int d \v{k} \, |\widetilde{\psi}_{\Lambda}(\v{k})|^2
\Gamma_{Y N}(\v{k})~,
\end{equation}
$\widetilde{\psi}_{\Lambda}(\v{p})$ denoting the Fourier transform of $\psi_{\Lambda}(\v{r})$.

Let us denote with $\v{r}$ the spatial point in which the final nucleon is
created and with $\v{r'}$ the spatial point in which the initial $\Lambda$
converts into the final $\Lambda$. Then, introduce a local nucleon
Fermi momentum depending on the position in which the final nucleon is created,
$k_{F}(r) = \{3 \pi^{2} \rho(r)/2\}^{1/3}$,
$\rho(r)$ being the density profile of the nuclear core.  It follows that the function
$\Gamma_{Y N}(\v{k},\v{r},\v{r'})$ is independent of $\v{r'}$ and can be written as
$\Gamma_{Y N}(\v{k},k_F(r))$. Finally, from Eqs.~(\ref{LDA0})--(\ref{LDA2})
one simply obtains Eq.~(\ref{decwYNld}), which formally is the same relation used
for the $\Lambda N\to nN$ non--mesonic decays.


\section{}
\label{APPEND}
Before presenting expressions for the evaluation of $\Gamma_{\Sigma^{0} n}$ and
$\Gamma_{\Sigma^{-} p}$, we call attention to some changes in
the baryon coupling constants with respect to our previous work.
As mentioned, the expressions for the functions
$S_{\tau}(q)$, $S'_{\tau}(q)$, $P_{L, \tau}(q)$,
$P_{C,\tau}(q)$, $P_{T, \tau}(q)$ and $S_{V, \tau}(q)$ appearing
in the weak transition potential $V^{\Lambda \Lambda \to Y N}$
of Eqs.~(\ref{intergral}) and (\ref{allmespot})
are given in Appendix B of Ref.~\cite{ba03}, where they refer to
the $V^{\Lambda N \to NN}$ potential.

The $V^{\Lambda \Lambda \to \Lambda n}$
transition potential, which is isoscalar, is obtained
by fixing $\tau=0$ in Eqs.~(\ref{intergral}) and (\ref{allmespot}) and by
making the following replacements for the strong coupling
constants:
$g_{NN \eta} \to g_{\Lambda \Lambda \eta}$,
$g^{V}_{NN \omega} \to g^{V}_{\Lambda \Lambda \omega}$,
$g^{T}_{NN \omega} \to g^{T}_{\Lambda \Lambda \omega}$.
Analogously, the $NN K$ and $NN K^{*}$ weak
parity conserving (PC) and parity
violating (PV) coupling constants are replaced by the
$\Lambda \Lambda K$ and $\Lambda \Lambda K^{*}$ couplings, respectively.
The two--pion--exchange weak potential has been obtained from the $\Lambda N\to \Lambda N$
scalar--isoscalar two--pion--exchange strong interaction potential
(including correlated and uncorrelated contributions) derived in
Ref.~\cite{Sa07} by a chiral unitary approach. This is obtained by
replacing the $g_{\pi NN}$ strong coupling constant
by the weak parity--conserving coupling $B_{\pi}=-7.15$
associated to the experimentally accessible $\Lambda N \pi$ vertex.

For the $V^{\Lambda \Lambda \to \Sigma^{0} n}$ transition potential,
which is isovector, we instead
fix $\tau=1$ in Eqs.~(\ref{intergral}) and (\ref{allmespot}). The relevant coupling
constants are obtained from the $V^{\Lambda N \to NN}$ ones by the following
replacements. For the strong coupling constants: $g_{NN \pi} \to g_{\Lambda \Sigma \pi}$,
$g^{V}_{NN \rho} \to g^{V}_{\Lambda \Sigma \rho}$ and
$g^{T}_{NN \rho} \to g^{T}_{\Lambda \Sigma \rho}$, while
for the weak coupling constants: $C^{PC}_{NN K} \to C^{PC}_{\Lambda \Sigma K}$,
$C^{PV}_{NN K} \to C^{PV}_{\Lambda \Sigma K}$,
$C^{PC}_{NN K^{*}} \to C^{PC}_{\Lambda \Sigma K^{*}}$ and
$C^{PV}_{NN K^{*}} \to C^{PV}_{\Lambda \Sigma K^{*}}$.

As explained in the text, by neglecting the small mass difference
between the hyperons $\Sigma^{0}$ and
$\Sigma^{-}$, isospin considerations lead to
$\Gamma_{\Sigma^{-} p}= 2 \, \Gamma_{\Sigma^{0} n}$.

After performing the summations over spin and isospin together with the
energy--integration one obtains the antisymmetrized $\Lambda \Lambda \to \Sigma^0 n$
decay rate in nuclear matter as:
\begin{eqnarray}
\label{gamdSn}
\Gamma_{\Sigma^{0} n}(\v{k},k_{F}) & =
& \frac{\pi}{3} \, (G_F m_{\pi}^2)^2  \int \frac{d^{3} p_1}{(2 \pi)^{3}} \int \frac{d^{3} p_2}{(2 \pi)^{3}}
\;  (2 \, {\cal W}^{dir}_{1}(q)- {\cal W}^{exch}(q,Q))\nonumber \\
& &  \times \theta(|\v{p}_2|-k_F) \;
\delta(k_0+k'_0-E_{\Sigma^{0}}(p_1)-E_{n}(p_2)),\nonumber
\end{eqnarray}
where $E_{\Lambda}$ ($E_{n}$) is the total $\Lambda$ (neutron) energy.
For the direct and exchange terms, the momentum matrix--element of the interaction
turn out to be:
\begin{equation}
\label{vdir2}
{\cal W}^{dir}_{1}(q)  =  \{
S^{2}_{1}(q) + S'^{2}_{1}(q)
 + P^{2}_{L, 1}(q)  +  P^{2}_{C, 1}(q) + 2  \, P^{2}_{T, 1}(q)+ 2 \, S^{2}_{V, 1}(q) \}\, ,
\end{equation}
and
\begin{eqnarray}
\label{vexc2}
{\cal W}^{exch}(q,Q) & = &
(\hat{\v{q}} \cdot \hat{\v{Q}})  \textsf{S}_{1}(q,Q)
+ ( 2 (\hat{\v{q}} \cdot \hat{\v{Q}})^2 - 1)  P_{L,1}(q)  P_{L, 1}(Q)  \nonumber \\
& & + 2 ((\hat{\v{q}} \cdot \hat{\v{Q}})^2 - 1)   P_{T,1}(q)  P_{T, 1}(Q) \nonumber \\
& &  -2 (\hat{\v{q}} \cdot \hat{\v{Q}})^2 (P_{L, 1}(q) P_{T, 1}(Q) + P_{L, 1}(Q) P_{T, 1}(q)) \nonumber \\
& & + P_{C, 1}(q) P_{C, 1}(Q) + P_{C, 1}(q) P_{L, 1}(Q) + P_{C, 1}(Q) P_{L, 1}(q) \nonumber \\
& & + 2 (P_{C, 1}(q) P_{T, 1}(Q) + P_{C, 1}(Q) P_{T,1}(q))\, ,
\end{eqnarray}
respectively, where $Q=q+k'-k$ and:
\begin{eqnarray}
\label{sexc2}
\textsf{S}_{1}(q,Q) & = &
(S_{1}(q) + S'_{1}(q))(S_{1}(Q)
+ S'_{1}(Q))
\nonumber \\
& & - 2 (S_{1}(q) S_{V, \, 1}(Q) +
S_{V, \, 1}(q) S_{1}(Q)) \nonumber \\
& & + 2 (S'_{1}(q) S_{V, \, 1}(Q) +
S_{V, \, 1}(q) S'_{1}(Q))\,.
\end{eqnarray}
The finite hypernucleus decay rate $\Gamma_{\Sigma^{0} n}$ is then obtained
by means of the LDA of Eq.~(\ref{decwYNld}).

\section*{Acknowledgements}
We would like to thank A. Ramos and A. Parre\~no for fruitful
discussions.
This work was partially supported by the
CONICET, Argentina, under contract PIP 0032 and
by the Agencia Nacional de Promociones Cient\'{\i}ficas
y T\'{e}cnicas, Argentina, under contract PICT-2010-2688,




\begin{thebibliography}{00}

\bibitem{reports}
E.~Botta, T.~Bressani and G.~Garbarino,
Eur. Phys. J. {\bf A 48} (2012) 41;
H.~Outa, in Hadron Physics (IOS Press, Amsterdam,
2005). Proc. of the International School of Physics 'E.
Fermi' Course CLVIII, edited by T. Bressani, A. Filippi
and U. Wiedner, p. 21;
W.M.~Alberico and G.~Garbarino,
Phys. Rep. {\bf 369} (2002) 1.

\bibitem{BG11}
E.~Bauer and G.~Garbarino,
Phys. Lett. {\bf B 698} (2011) 306.

\bibitem{Nak10}
K.~Nakazawa and H.~Takahashi,
Prog. Theor. Phys. Suppl. {\bf 185} (2010) 335.

\bibitem{Tak01}
J. K. Ahn {\it et al.},
Phys. Rev. {\bf C 88} (2013) 014003;
H.~Takahashi {\it et al.}
Phys. Rev. Lett. {\bf 87} (2001) 212502.

\bibitem{GM12}
A. Gal and D.J. Millener,
Phys. Lett. {\bf B 701} (2011) 342;
Hyperfine Interact. {\bf 210} (2012) 77.

\bibitem{J-PARC}
T. Takahashi,
Nucl. Phys. {\bf A 914} (2013) 530.

\bibitem{Panda}
A. Sanchez Lorente {\it et al.,}
Hyperfine Interact. {\bf 229} (2014) 45;
U. Wiedner,
Prog. Part. Nucl. Phys. {\bf 66} (2011) 477.

\bibitem{Wa07}
T.~Watanabe {\it et al.},
Eur. Phys. J. {\bf A 33} (2007) 265.


\bibitem{Pa01}
A. Parre\~{n}o, A. Ramos and C. Bennhold,
Phys. Rev. {\bf C 65} (2001) 015205.

\bibitem{It01}
K. Itonaga, T. Ueda and T. Motoba,
Nucl. Phys. {\bf A 691} (2001) 197c.
Mod. Phys. Lett. {\bf A 18} (2003) 135;


\bibitem{Sa03}
K. Sasaki, T. Inoue and M. Oka,
Nucl. Phys. {\bf A 726} (2003) 349.



\bibitem{BG09}
E.~Bauer and G.~Garbarino,
Nucl. Phys. {\bf A 828} (2009) 29.

\bibitem{BG10}
E.~Bauer and G.~Garbarino,
Phys. Rev. {\bf C 81} (2010) 064315.

\bibitem{BGPR12}	
E. Bauer, G. Garbarino, A. Parre\~{n}o and A. Ramos,
Phys.Rev. {\bf C 85} (2012) 024321.

\bibitem{BG12}
E.~Bauer and G.~Garbarino,
Phys. Lett. {\bf B 716} (2012) 249.

\bibitem{Os85}
E. Oset and L. L. Salcedo, Nucl. Phys. {\bf A 443}, 704 (1985).

\bibitem{Ca99}
J. Caro, C. Garcia--Recio and J. Nieves,
Nucl. Phys. {\bf A 646} (1999) 299.

\bibitem {ba03}
E. Bauer and F. Krmpoti\'c, Nucl. Phys. {\bf A 717} (2003) 217.

\bibitem{Sa07}
K. Sasaki, E. Oset, and M.J. Vicente Vacas,
Phys. Rev. {\bf C 74} (2006) 064002;
K. Sasaki, E. Oset and M. J. Vicente Vacas,
Eur. Phys. J. {\bf A 31} (2007) 557.

\bibitem{It02}
K. Itonaga,T. Ueda and T.Motoba,
Phys. Rev. {\bf C 65} (2002) 034617.

\bibitem{Do86}
J. F. Donoghue, E. Golowich and B. R. Holstein,
Phys.Rev. {\bf D 34} (1986) 3434.

\end{thebibliography}
\end{document}